\documentclass[a4paper,11pt]{article}
\pdfoutput=1 

\usepackage{jcappub}
\usepackage[T1]{fontenc} 

\makeatletter
\gdef\@fpheader{}
\makeatother

\usepackage{graphicx}
\usepackage{color}
\usepackage[colorlinks=true, urlcolor=blue, linkcolor=blue, 
citecolor=blue]{hyperref}

\newcommand{\dd}{{\rm d}}
\newcommand{\N}{{\rm N}}
\newcommand{\Ox}{{\rm O}}

\newcommand{\Xm}{X_{\rm max}}
\newcommand{\beq}{\begin{equation}}
\newcommand{\eeq}{\end{equation}}
\newcommand{\beqa}{\begin{eqnarray}}
\newcommand{\eeqa}{\end{eqnarray}}
\newcommand{\beqar}{\begin{eqnarray*}}
\newcommand{\eeqar}{\end{eqnarray*}}

\begin{tiny}

\end{tiny} 

\title{\boldmath Ultra-peripheral collisions of charged hadrons in extensive air showers}

\author[a]{Manuel Masip,}
\author[a]{Ivan Rosario,}
\author[b]{Sergio J. Sciutto}

\affiliation[a]{CAFPE and Departamento de F{\'\i}sica Te\'orica y del Cosmos, Universidad de Granada, E-18071 Granada, Spain}
\affiliation[b]{Departamento de F{\'\i}sica e Instituto de F{\'\i}sica La Plata (CONICET), Universidad Nacional de La Plata, C.C. 69 (1900) La Plata, Argentina}

\emailAdd{masip@ugr.es}
\emailAdd{ivan.rosario@ugr.es}
\emailAdd{sciutto@fisica.unlp.edu.ar}

\abstract{We discuss the electromagnetic collisions of high energy protons, pions and kaons with 
atmospheric nuclei. In particular, we use the equivalent photon approximation to 
estimate {\it (i)} the diffractive collisions where the projectile scatters inelastically off a 
nucleus, and {\it (ii)} the usual radiative processes (bremsstrahlung, pair production and
photonuclear interactions) of these charged hadrons in the air. 
We then include the processes in the simulator AIRES and study how they affect the
longitudinal development of extensive air showers. 
For $10^{9-11}$ GeV proton primaries we find that they introduce a very small reduction
(below 1\%) in the average value of both $\Xm$ and $\Delta \Xm$. At a given shower 
age (relative slant depth from
$\Xm$), these electromagnetic processes slightly increase the number of charged particles
at the shower maximum and reduce the number of muons when it is old, decreasing by 1\%
the muon-to-($\gamma+e$) near the ground level.}

\begin{document}

\maketitle
\flushbottom

\section{Introduction} 

Cosmic rays (CRs), with energies of up to $10^{11}$ GeV, provide a window for the exploration of collisions
at extreme energies. When they reach the Earth, the atmosphere acts like a calorimeter of very low density 
but equivalent 
to 10 m of water\footnote{Or 20 m.w.e. if the primary enters from a zenith inclination $\theta_z=60^\circ$.}, 
resulting in an extensive air shower (EAS) that includes
three basic components: a hadronic one, an electromagnetic (EM) one with photons and electrons, plus  a component with muons
and neutrinos from light meson decays \cite{Gaisser:1990vg, Lipari:1993hd}. 
Fluorescence and surface detectors at observatories like AUGER \cite{PierreAuger:2016qzd}
can then estimate the total energy of the primary, the atmospheric depth $\Xm$ with 
the maximum number of charged particles, and the number and distribution of electrons and muons reaching the ground.
The relation of these observables with the spectrum and composition of the primary CR flux faces an obvious difficulty:
since we do not have access to a controlled source of CRs, this atmospheric calorimeter can not be properly 
calibrated, and the results will heavily rely on Monte Carlo simulations.

\begin{figure}
\vspace{-0.5cm}
\begin{center}
\includegraphics[width=0.5\textwidth]{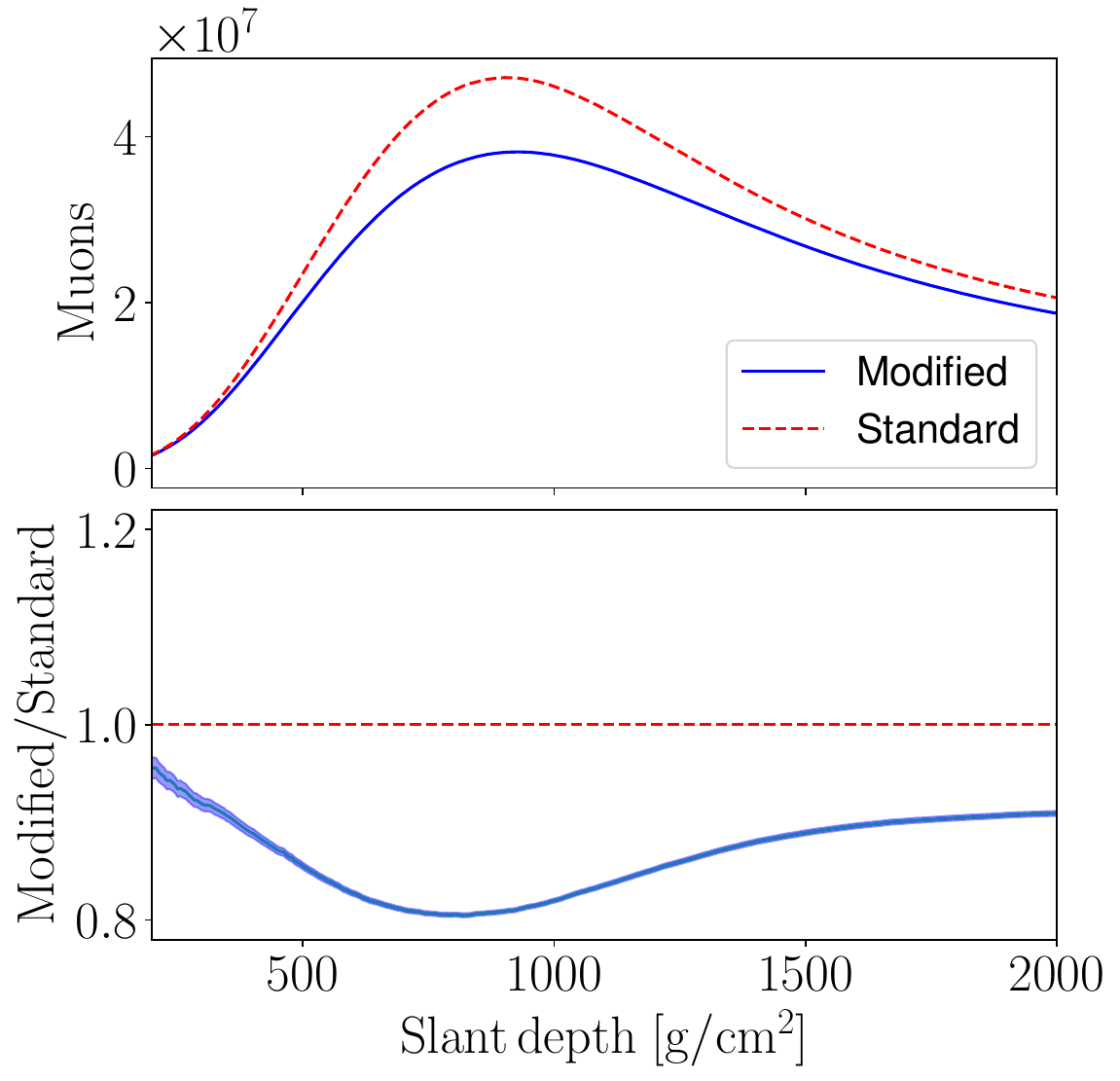}
\end{center}
\vspace{-0.5cm}
\caption{Muon content in the average $10^{10}$ GeV proton shower with (dashes) or without (solid) hadronic interactions of 
photons (from 40000 simulations with AIRES, $\theta_z=70^\circ$). The deficit (lower figure) goes from 20\% at $\Xm$ to 8\% at
the surface.
\label{f1}}
\end{figure}

It then becomes essential to identify all the relevant processes in the EAS 
and the main sources of uncertainty in the simulations \cite{Ulrich:2009hm}.
In particular, CR collisions involve a high energy regime\footnote{Notice that a $10^8$ GeV proton hitting an atmospheric
nucleon reproduces the 14 TeV center of mass energy currently studied at the LHC.} and a kinematical region (ultraforward
rapidities are critical in the longitudinal development of a shower) that are of difficult access at colliders.
Moreover, lower energy processes may introduce corrections that, given the large number of collisions  
in the core of an EAS before it reaches the ground, may become sizeable.
For example, most of the EM energy in the shower is generated through $\pi^0$ decays 
high in the atmosphere. As this energy goes forward it degrades after each
radiation length $X_0\approx 38$ g/cm$^2$. Since the cross section for a hadronic collision of a
photon with an air nucleus 100 times smaller, this degradation happens mostly through purely EM processes ($\gamma\,A \to e^+ e^-A$ and $e\, A\to e\gamma\,A$). This initial EM energy, however, crosses a large depth before it is
absorbed ({\it e.g.}, it takes around 20 radiation lengths to reduce a $10^6$ GeV photon to 1 GeV electrons and photons). As a consequence, the probability that part of the EM energy goes back to hadrons
within that interval can not be ignored. 
This is illustrated in Fig.~\ref{f1} for $10^{10}$ GeV proton showers.
If the hadronic collisions of photons are turned off in AIRES \cite{Sciutto:1999jh}, we find an $8.3\%$ reduction in 
the average number of muons from meson decays at the ground level.

Here we will discuss some effects that are neglected in current EAS simulators like AIRES or CORSIKA \cite{Engel:2018akg,Heck:1998vt}: the photon-mediated
ultra-peripheral collisions of charged hadrons with atmospheric nuclei. 
A simple argument suggests that these processes should be taken into account. 
When one of these hadrons crosses the EM field of a nucleus it may get diffracted into 
a system of mass $m^* > m + m_\pi$ giving a final state with several
hadrons, {\it e.g.}, 
\beq
p \, A \to N\, \pi\, A\;.
\eeq
Notice that at higher projectile energies, this inelastic 
process may occur at larger transverse distances: unlike the pomeron-mediated 
diffractive cross section, this one grows with the energy \cite{Guzey:2005tk}. Its possible relevance may remind us to what happens in the propagation of CRs through the intergalactic medium, where the collisions with the
CMB photons are irrelevant until they become inelastic at the GZK \cite{Greisen:1966jv, Zatsepin:1966jv} energy. 

We will also discuss the radiative emissions of the charged hadrons in the atmosphere, namely, bremsstrahlung
(BR), pair production (PP) and photonuclear (PN) collisions where the projectile is still present after the collision:
\beq
h\, A \to h\,\gamma\, A\;; \quad
h\, A \to h\,e^+e^- A\;; \quad
h\, A \to h\,\rho\, A \to h\, X\,.
\eeq
with $h=p,K,\pi$. These processes are ultra-peripheral as well, at impact parameters $b>R_A$, with the EM 
field of $h$ going into an $e^+ e^-$ pair (PP) or a $\rho$ meson (PN) or with the projectile scattering off
the EM field of the nucleus and emitting a photon (BR). We will use the equivalent photon approximation (EPA) to estimate 
the rate of all these processes and will discuss the validity of this approximation.
Our objective is to obtain the precise effect of ultra-peripheral EM collisions
in the longitudinal development of EASs.

\section{Bremsstrahlung and diffractive collisions} 

A relativistic charged particle creates an EM field that can be approximated by a cloud of virtual photons \cite{Fermi:1925fq}. 
These photons 
may interact with the photon 
cloud of another charged particle (in a $\gamma \gamma$ collision) 
or with the target particle itself. Notice that if the transverse distance between two charged hadrons in a collision is 
$b>R_1+R_2$, these ultra-peripheral processes will not occur simultaneously with a hadronic one. For an atmospheric
nitrogen nucleus $\N$, the equivalent photons are coherently radiated, which imposes a limit on their minimum wavelength.

Let us be more specific. Consider a hadron $h$ of energy $E$ and mass $m_h$ moving in the atmosphere. In its rest frame, 
$h$ sees the nucleus $\N$ approaching with
a Lorentz factor $\gamma=E/m_h$ and surrounded by the cloud of photons. In the transverse plane the photons have a
momentum $p_T\le 1/R_\N\approx 71$ MeV, whereas in the longitudinal direction their momentum can be much larger,
$p_L\le \gamma/R_\N$. The virtuality, $|q^2|<1/R_N^2$, of these quasi-real photons is small compared to their energy.
The total flux of equivalent photons around the nucleus 
is obtained with the Weizsaker-Williams method; upon integration in impact parameter
space between $b_{\rm min}$ and $b_{\rm max}$ it gives \cite{Bertulani:2005ru}
\beqa\label{eq:photonSpectrum}
{\dd N_\gamma \over \dd \omega}  = {\alpha \,Z^2 \over \pi\, \gamma^2} \left[
\omega \, b^2 \left( K_0(x)^2 -  K_1(x)^2 \right)+ 2 \gamma \, b  K_1(x)\,K_0(x)\right] \Bigg|_{b_{\rm min}}^{b_{\rm max}} ,
\eeqa
where $\omega$ is the energy of the photons, 
$K_n(x)$ modified Bessel functions of the second kind, $ x={\omega b/ \gamma}$,  
$b_{\rm min}=R_{\N}$  and $b_{\rm max} \approx 1/( \alpha \,m_e)$. For the radius of a
nucleus we will take $R_A=5.8\,A^{1/3}$ GeV$^{-1}$.

\begin{figure}
\begin{center}
\includegraphics[width=0.5\textwidth]{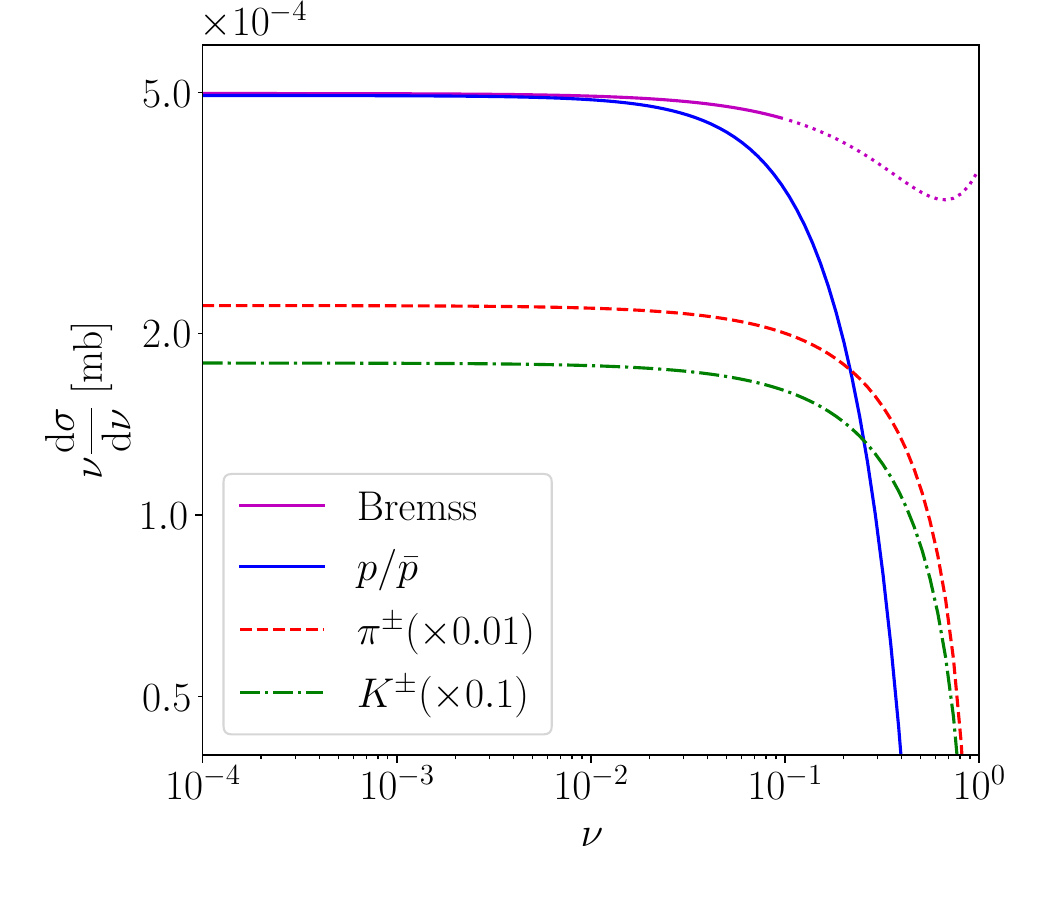}
\end{center}
\vspace{-0.5cm}
\caption{Bremsstrahlung cross section off a nitrogen nucleus for a point-like proton \cite{Groom:2001kq} and our estimate obtained 
using the equivalent
photon approximation for charged hadrons.
\label{f2}}
\end{figure}

It is then easy to describe the collision of this equivalent photon flux with a hadron $h$ at rest. At low values of 
$\omega$ ($\omega\le 1$ GeV) the dominant process is just Compton scattering; for $h=p$
 the differential cross section reads
\beq
{\dd \sigma_{\gamma p \to \gamma p} \over \dd \cos\theta} ={\pi \alpha^2\, |F(t)|^2
\over m_p^2} \left( {\omega'\over \omega} \right)^2 \left(
{\omega'\over \omega} +{\omega \over \omega'} -1+\cos^2\theta \right),
\eeq
where $\theta$ is the scattering angle and $\omega'=\omega \left( 1+{\omega\over m_p}\left(1-\cos\theta\right)\right)^{-1}$ is the energy of the final photon. In the expression above we have included a form factor
\beq
F(t)={m_p^2-0.7\,t\over \left(m_p^2-0.25\, t\right) \left(1-\displaystyle{t\over (0.7\,{\rm GeV})^2}\right)^2}
\eeq
that suppresses elastic scatterings with large momentum transfer.
Going back to the frame with the nucleus at rest, we can express this cross section in
terms of the fraction of energy $\nu$ lost by the incident proton\footnote{Notice that $\nu$ and 
$\omega$ are kinematical variables defined in different reference frames.}:
\beq
{\dd \sigma_{p\gamma \to p\gamma} \over \dd \nu} =
{\pi \alpha^2\, |F(t)|^2  \over m_p\,\omega} \left( {1-\nu+\nu^2\over 1-\nu} + \left( 1- {m_p\over \omega}
\,{\nu\over 1-\nu} \right)^2 \right),
\eeq
with $t=-2\omega\nu m_p$.
Adding the contribution of all the equivalent photons we obtain
\beq
{\dd \sigma_{p\N\to p\gamma \N} \over \dd \nu} = \int 
\dd \omega\,{\dd N_\gamma \over \dd \omega}   \, 
{\dd \sigma_{p\gamma \to p\gamma} \over \dd \nu}\,,
\eeq
with $\omega_{\rm min}={m\over 2}\,{\nu\over 1-\nu}$. 
We find that this cross section for bremsstrahlung ($p \N \to p \gamma \N$) obtained using 
inverse Compton scattering ($p\gamma \to p \gamma$, where
the $\gamma$ is an equivalent photon around the nitrogen nucleus) gives an excellent 
approximation to the explicit calculation (see \cite{Groom:2001kq} and references therein). 
In Fig.~\ref{f2} we compare both cross sections
for a $10^{10}$ GeV proton (the dependence with the energy of the projectile for $E>1$ TeV is negligible).
In the EPA (see Eq.~(\ref{eq:photonSpectrum})) 
we have taken $b_{\rm max} =\pi/( \alpha \,m_e)$, whereas  the bremsstrahlung cross section corresponds
to a point-like proton (the form factor suppresses the differential cross section only at $\nu\ge 0.1$).
An analogous calculation for charged mesons ($h=\pi,K$), with
\beq
{\dd \sigma_{h\gamma \to h\gamma} \over \dd \nu} =
{\pi \alpha^2\, |F(t)|^2  \over m_h\,\omega} \left( 1 + \left( 1- {m\over \omega}
\,{\nu\over 1-\nu} \right)^2 \right),
\eeq
gives the cross sections also included in Fig.~\ref{f2}.

\begin{figure}[!t]
\vspace{-0.5cm}
\begin{center}
\includegraphics[width=0.49\textwidth]{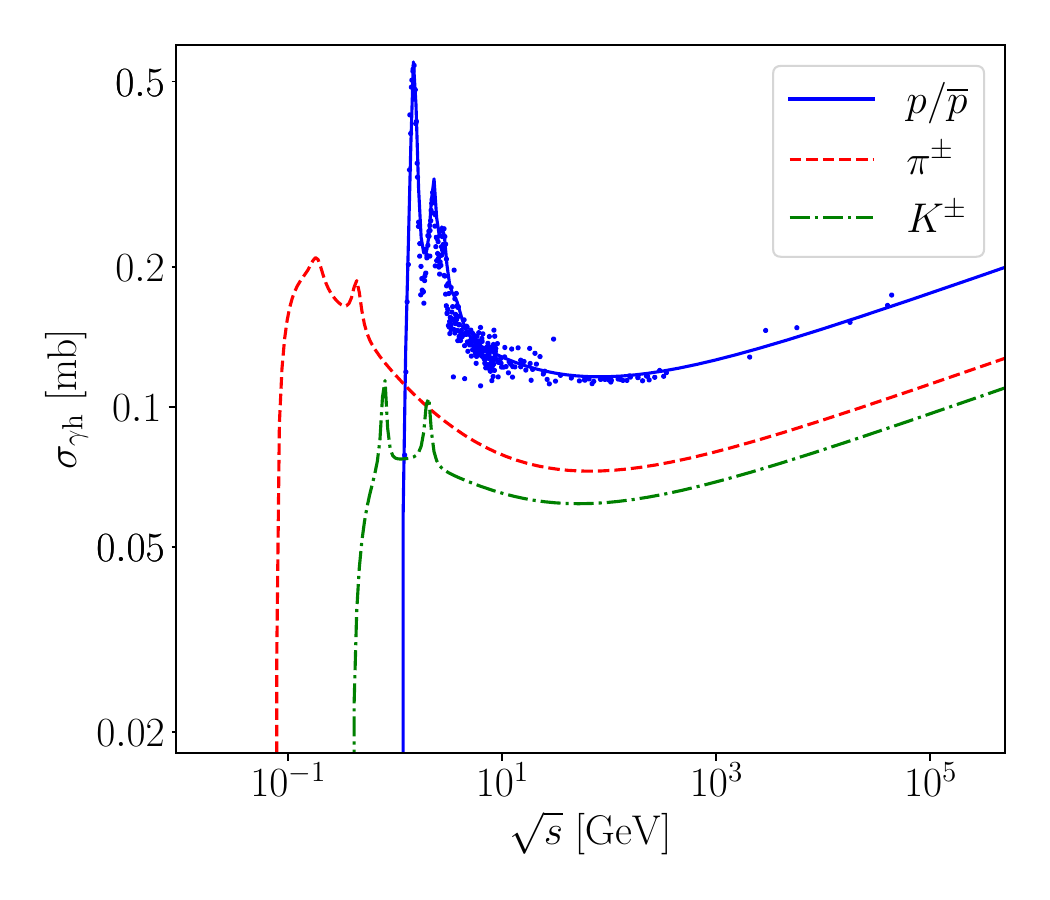}
\includegraphics[width=0.495\textwidth]{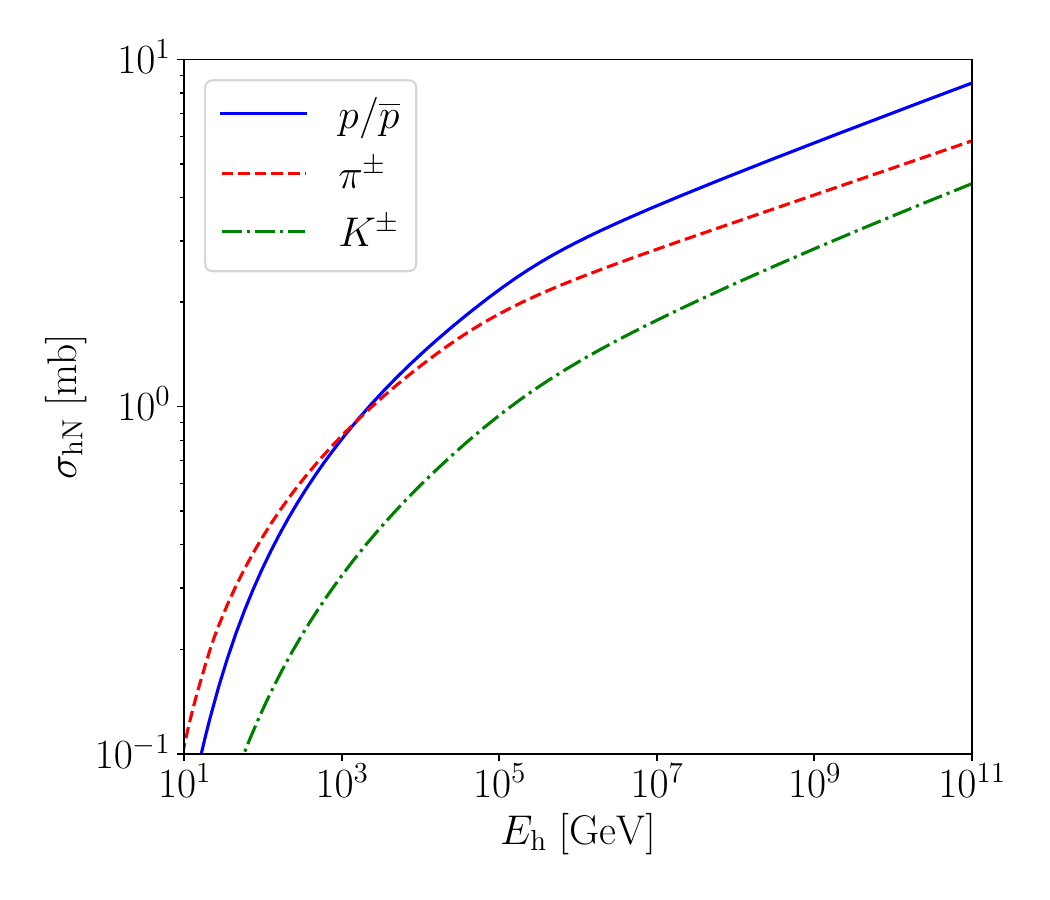}
\end{center}
\vspace{-0.5cm}
\caption{Cross section $\sigma_{\gamma h}$ and our estimate for the diffractive cross section 
$\sigma_{h \N}^{\rm diff}$ in the EPA.
\label{f3}}
\end{figure}

We can now estimate the collision of the projectile $h$ with equivalent photons of higher energy, 
$\omega\ge 1$ GeV in the frame with $h$ at rest. These are inelastic collisions ($\gamma h\to X$) 
where the hadron absorbs the photon and goes to a final state with pions \cite{Mucke:1999yb}. 
In Fig.~\ref{f3}-left we plot our fit for such collisions; we include the first resonances 
 plus 
\beq
\sigma_{\gamma h}(s)= A_h \, s^{0.0808} + B_h \,s^{-0.4525}\,,
\eeq
with $A_p=0.069$, $B_p=0.129$; $A_\pi=0.044$, $B_\pi=0.0734$; $A_K=0.038$, $B_p=0.059$ and 
$s=2\omega m_h+m_h^2$. Adding the contribution
of all the photons in the N cloud we obtain the total diffractive cross section in
Fig.~\ref{f3}-right. In these EM processes 30\% of the cross section comes from collisions with low-energy equivalent 
photons that take the incident 
projectile to a hadronic resonance: $\Delta(1232)$ to $\Delta(1950)$ for the proton, $\rho(770)$ to $a_2(1320)$ for pions,
and $K^*(892)$ to $K_2^*(1430)$ for kaons. In this case, the final state will typically include an extra pion carrying a fraction
$m_\pi/(m_\pi+m_h)$ of the incident energy, whereas 
in the remaining $70\%$ of the cases the final state will include several pions.

The EM diffractive cross section that we have obtained implies an interaction length (in g/cm$^2$) in nitrogen 
$\lambda_{h\N}^{\rm diff}=m_\N/\sigma_{h \N}^{\rm diff}$. In the air, if we take a $72\%$ N plus $28\%$ O composition,
\beq
{1\over \lambda_{h\N}^{\rm diff}}={0.72\,\sigma_{h \N}^{\rm diff} \over m_\N}+{0.28\,\sigma_{h \Ox}^{\rm diff} \over m_\Ox}\,.
\eeq
Since $\sigma_{h \Ox}^{\rm diff}/\sigma_{h \N}^{\rm diff}\approx (8/7)^2$, we obtain
\beq
\lambda_{h{\rm \,air}}^{\rm diff}\approx 0.96\; {m_\N \over \sigma_{h \N}^{\rm diff}}\,,
\eeq
with a 69\% probability for a $h \N$ collision  and a 31\% probability for a collision with O. These approximate
relations hold for bremsstrahlung and pair production as well.

\section{Pair production and photonuclear collisions} 

The two processes discussed in the previous section can be understood as the collision of the projectile with the
photon cloud around the
atmospheric nucleus. But we also have the opposite process, the collision of the equivalent photons carried by
the charged hadron with the nucleus. 
Obviously, these collisions will only depend on the velocity (or the Lorentz factor $\gamma=E/m_h$) of $h$.
In the frame with the nucleus at rest, their spectrum is given by 
\beq
{\dd N_\gamma \over \dd \omega}  = {\alpha \,b_{\rm min}\over \pi\, \gamma^2} \left(
\omega \, b_{\rm min}\, K_0(x_{\rm min})^2+ 2 \gamma \, K_1(x_{\rm min})\,K_0(x_{\rm min})- 
\omega \, b_{\rm min}\, K_1(x_{\rm min})^2  \right),
\eeq
with $x_{\rm min}=\omega b_{\rm min}/\gamma$ and $b_{\rm min}=(0.17\,{\rm GeV})^{-1}$. As the equivalent photons propagate 
in the atmosphere they
may create an $e^+ e^-$ pair or experience a photonuclear collision. 
Let us first discuss pair production \cite{Klein:2004is}.

\begin{figure}
\begin{center}
\includegraphics[width=0.5\textwidth]{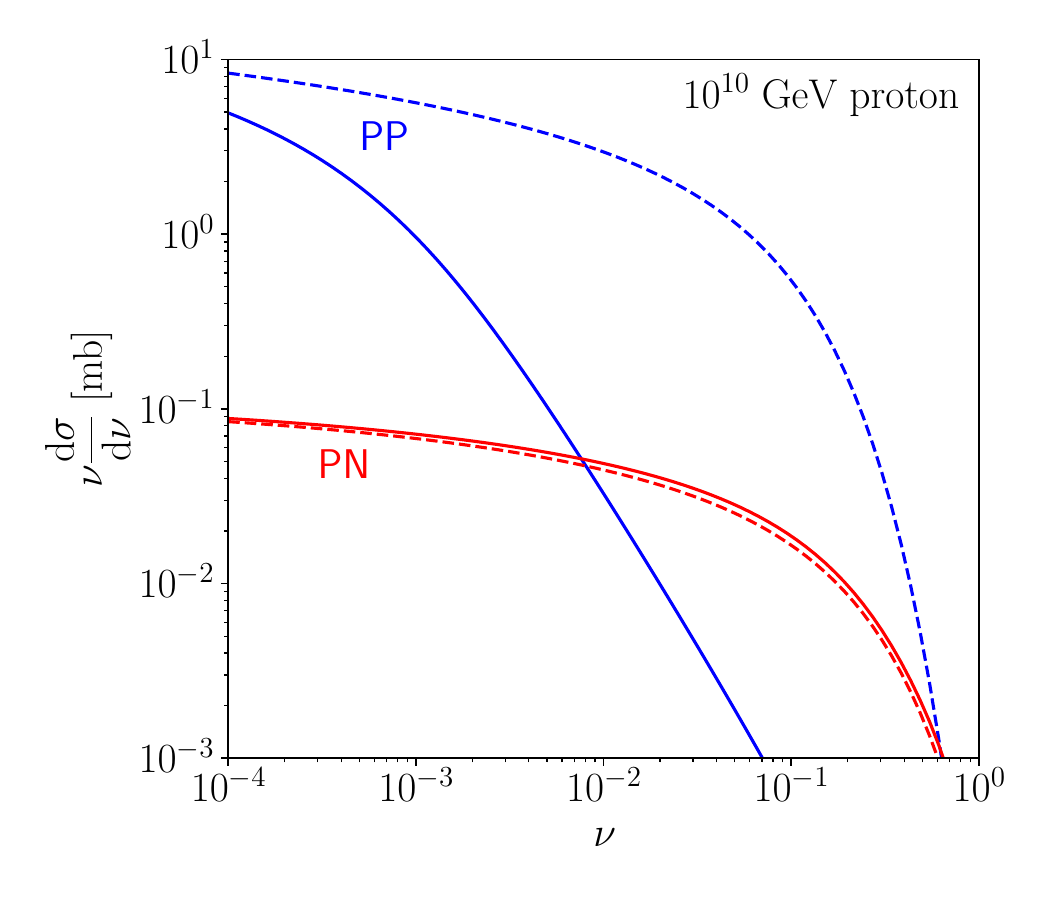}
\end{center}
\vspace{-0.5cm}
\caption{Pair production (PP) and photonuclear (PN) cross sections obtained using the 
equivalent photon approximation (dashes) and from an explicit calculation (solid).
\label{f4}}
\end{figure}

At photon energies above 10 GeV the cross section to convert into a pair becomes constant,
\beq
\sigma_\gamma = {7\,m_A\over 9\,X_0}\,,
\eeq
where $m_A$ is the mass of the nucleus in the medium and $X_0$ the radiation length. Including  screening, 
 collisions with electrons, and radiative corrections one has
\beq
X_0={m_A\over 4\alpha r_e^2 \left( Z^2 \left( \ln {184\over Z^{1/3}} - f(Z) \right) + Z \ln {1194\over Z^{2/3}} \right)}\,,
\eeq
with 
\beq
f(Z)=\left( \alpha Z \right)^2\, \sum_{n=1}^\infty{1\over n\left( n^2 + \left( \alpha Z \right)^2 \right) }\,.
\eeq
In a nitrogen medium $\sigma_\gamma=470$ mb and $X_0=38.4$ g/cm$^2$, whereas in the atmosphere the radiation length
is a 4\% shorter (see the discussion at the end of Section 2). Including all the photons in the cloud, 
the differential cross section for a hadron $h$ of
energy $E$ to loose a fraction $\nu$ of its energy by the conversion of one of these photons 
into an $e^+ e^-$ pair would be
\beq
{\dd \sigma({h\N \to h e^+e^-\N)}\over \dd \nu}\equiv {\dd \sigma_h\over \dd \nu}
= E \, {\dd N_\gamma \over \dd \omega} \,\sigma_\gamma\,.
\eeq
In Fig.~\ref{f4}, we plot this result for a proton projectile (dashes)
together with the result from an explicit calculation (solid) \cite{Groom:2001kq}. 
The EPA overestimates the pair production cross section; 
indeed, to neglect the off-shellness of these
equivalent photons is not a good approximation when the final state has an invariant mass of order $2m_e$ \cite{Budnev:1975poe}. 
For $E>1$ TeV, this cross
section is independent from the projectile energy. The differential cross section for pions and kaons can be 
readily obtained from
\beq
\left.{\dd \sigma_h\over \dd \nu}\right|_{(\nu,E)}= r_h \left.{\dd \sigma_p\over \dd \nu}\right|_{(r_h \nu ,\, r_h^{-1}E)}
\label{rel}
\eeq
where $r_h\equiv {m_h/m_p}$,

Finally, there are the processes where the photons in the cloud around $h$ experience a hadronic collision
(radiative photonuclear collisions \cite{Schuler:1993td}). Assuming vector
meson dominance, a photon carrying a fraction $\nu$ of the hadron's energy fluctuates
into a $q\bar q$ pair; the pair forms a $\rho$ (or a $J/\Psi$) meson that may then interact elastically
($\gamma N \to \rho N$) or inelastically with the nitrogen nucleus. In Fig.~\ref{f4}, we include the result for a
$10^{10}$ proton projectile (the relation in Eq.~(\ref{rel}) for pions and
kaons is also valid in this case) together
with the explicit calculation of the cross section \cite{Borog:1975yq, Brun:1978fy}. 
We see that in this case the EPA gives an excellent agreement. 
We provide fits for all these ultra-peripheral 
processes in the next section.

\section{AIRES simulations} 

AIRES includes the energy loss by ionization of charged hadrons, but not the radiative emissions
considered in previous sections. These dominate over ionization 
at Lorentz factors above $\gamma_c\approx 2000$, {\it i.e.}, 
at energies above 2 TeV for protons, 1 TeV for kaons and 300 GeV for pions. Our objective is then to modify the propagation
of these high energy charged hadrons in an EAS.

In Table 1, we provide the different interaction lengths in air for protons, pions and kaons at energies between $10^2$ and
$10^{11}$ GeV. The lengths $\lambda_{\rm BR}^h$ and $\lambda_{\rm PP}^h$ correspond to energy depositions
$E_{\rm dep}>0.1$ GeV or $\nu_{\rm min}={0.1\,{\rm GeV}\over E}$, whereas in photonuclear depositions we take
\beq
{E_{\rm dep}\over \;{\rm GeV}}> \sqrt{E \over 100\;{\rm GeV}}\;\;\; {\rm or}\;\;\; \nu_{\rm min}= 0.1 \sqrt{{\rm GeV}\over E}\,.
\eeq
Up to an energy-dependent normalization, the approximate $\nu$ distribution in each process is the following.
In a bremsstrahlung collision
\beq
f_{\rm BR}^\pi(\nu)={1\over \nu}\, \left( 1-\nu \right)^{1.25} \,;
\eeq
\beq
f_{\rm BR}^K(\nu)={1\over \nu}\, \left( 1-1.2\;\nu \right)^{1.25} \,,
\eeq
\beq
f_{\rm BR}^p(\nu)={1\over \nu}\, \left( 1-1.2\,\nu \right)^{2.20} \,;
\eeq
with $\nu\le 0.8$. In the emission of an $e^+ e^-$ pair by a pion projectile the $\nu$ distribution is given by
\beq
f_{\rm PP}^\pi(\nu)={ 1-\nu \over \nu^{1.18}\left( 1+4571\, \nu^{2.64} \right) } \,,
\eeq
whereas for $h=p,K$
\beq
f_{\rm PP}^h(\nu) = {m_h\over m_\pi}\, f_{\rm PP}^\pi(m_h\,\nu/m_\pi)\,.
\label{rel2}
\eeq
Finally, in a photonuclear collision we obtain
\beq
f_{\rm PN}^\pi(\nu)={ 1-\nu^{0.22} \over \nu^{0.981} } \,,
\eeq
with a negligible dependence on the energy of the projectile (other than the dependence in $\nu_{\rm min}$).
The distribution for protons and kaons given also by the relation in (\ref{rel2}).

\begin{table}[!t]
{\small
\begin{center}
\begin{tabular}{ c|ccccc| } 
 \hline
 \hline
Energy [GeV] & $\lambda_{\rm had}^p$ [g/cm$^2$] & $\lambda_{\rm BR}^p$  [g/cm$^2$] & $\lambda_{\rm DIFF}^p$ [g/cm$^2$]  & $\lambda_{\rm PP}^p$  [g/cm$^2$]  & $\lambda_{\rm PH}^p$ [g/cm$^2$]   \\ 
 \hline
$10^3$& $83.3$ & $76.8\times 10^5$ & $26.2\times 10^3$ & $33.1\times 10^2$ & $31.9\times 10^4$ \\
$10^4$& $76.7$ & $50.0\times 10^5$ & $15.2\times 10^3$ & $940$ & $20.6\times 10^4$ \\
$10^5$& $71.1$ & $38.5\times 10^5$ & $99.5\times 10^2$ & $426$ & $13.8\times 10^4$ \\
$10^6$& $64.1$ & $32.1\times 10^5$ & $73.0\times 10^2$ & $241$ & $93.1\times 10^3$ \\
$10^7$& $56.6$ & $27.6\times 10^5$ & $57.9\times 10^2$ & $155$ & $63.7\times 10^3$ \\
$10^8$& $50.5$ & $24.3\times 10^5$ & $47.0\times 10^2$ & $108$ & $44.4\times 10^3$ \\
$10^9$& $45.5$ & $21.7\times 10^5$ & $38.4\times 10^2$ & $79.1$ & $31.7\times 10^3$ \\
$10^{10}$& $41.6$ & $19.5\times 10^5$ & $31.5\times 10^2$ & $60.5$ & $23.1\times 10^3$ \\
$10^{11}$& $38.3$ & $17.8\times 10^5$ & $25.9\times 10^2$ & $48.1$ & $19.4\times 10^3$ \\
 \hline
 \hline
Energy [GeV] & $\lambda_{\rm had}^\pi$ [g/cm$^2$] & $\lambda_{\rm BR}^\pi$  [g/cm$^2$] & $\lambda_{\rm DIFF}^\pi$ [g/cm$^2$]  & $\lambda_{\rm PP}^\pi$  [g/cm$^2$]  & $\lambda_{\rm PH}^\pi$ [g/cm$^2$]   \\ 
 \hline
$10^3$& $111$ & $13.3\times 10^4$ & $26.3\times 10^3$ & $11.2\times 10^2$ & $18.3\times 10^4$ \\
$10^4$& $99.5$ & $99.7\times 10^3$ & $16.6\times 10^3$ & $480$ & $12.7\times 10^4$ \\
$10^5$& $89.3$ & $81.0\times 10^3$ & $11.7\times 10^3$ & $264$ & $89.1\times 10^3$ \\
$10^6$& $79.8$ & $68.5\times 10^3$ & $92.3\times 10^2$ & $166$ & $62.5\times 10^3$ \\
$10^7$& $69.3$ & $59.4\times 10^3$ & $76.4\times 10^2$ & $114$ & $44.3\times 10^3$ \\
$10^8$& $59.3$ & $52.5\times 10^3$ & $63.9\times 10^2$ & $83.3$ & $31.8\times 10^3$ \\
$10^9$& $52.2$ & $46.8\times 10^3$ & $53.6\times 10^2$ & $63.4$ & $23.3\times 10^3$ \\
$10^{10}$& $46.9$ & $42.5\times 10^3$ & $44.9\times 10^2$ & $49.7$ & $17.4\times 10^3$ \\
$10^{11}$& $42.9$ & $38.7\times 10^3$ & $37.5\times 10^2$ & $40.4$ & $14.7\times 10^3$ \\
 \hline
 \hline
Energy [GeV] & $\lambda_{\rm had}^K$ [g/cm$^2$] & $\lambda_{\rm BR}^K$  [g/cm$^2$] & $\lambda_{\rm DIFF}^K$ [g/cm$^2$]  & $\lambda_{\rm PP}^K$  [g/cm$^2$]  & $\lambda_{\rm PH}^K$ [g/cm$^2$]   \\ 
 \hline
$10^3$& $125$ & $18.9\times 10^5$ & $60.7\times 10^3$ & $21.6\times 10^2$ & $25.6\times 10^4$ \\
$10^4$& $115$ & $13.1\times 10^5$ & $34.1\times 10^3$ & $730$ & $17.1\times 10^4$ \\
$10^5$& $103$ & $10.3\times 10^5$ & $21.8\times 10^3$ & $357$ & $11.7\times 10^4$ \\
$10^6$& $89.9$ & $86.9\times 10^4$ & $15.6\times 10^3$ & $211$ & $80.0\times 10^3$ \\
$10^7$& $76.7$ & $75.1\times 10^4$ & $12.0\times 10^3$ & $139$ & $55.6\times 10^3$ \\
$10^8$& $65.1$ & $66.0\times 10^4$ & $94.8\times 10^2$ & $98.4$ & $39.2\times 10^3$ \\
$10^9$& $57.0$ & $59.1\times 10^4$ & $76.0\times 10^2$ & $73.2$ & $28.3\times 10^3$ \\
$10^{10}$& $51.1$ & $53.4\times 10^4$ & $61.4\times 10^2$ & $56.5$ & $20.8\times 10^3$ \\
$10^{11}$& $46.6$ & $48.7\times 10^4$ & $49.9\times 10^2$ & $45.3$ & $17.5\times 10^3$ \\
 \hline
 \hline
\end{tabular}
\caption{Interaction length in air for the different processes, projectiles and energies.}
\end{center}
}
\end{table}

The implementation of these processes in AIRES has been done in two steps: {\it (i)} we modify the mean free path (shortened by the new interactions) and find the relative frequency of each process, and {\it (ii)} we characterize the final state for these processes. 

In bremsstrahlung and pair-production the final state includes a real photon or an $e^+ e^-$ pair with the $\nu$-distributions given above. In our estimate we will take all the radiative emissions in the direction of the projectile, with equal energy for the two electrons in the pair. 
In a diffractive collision the energy of the equivalent photon around the nucleus is sampled. We will assume a final state with a leading hadron (a nucleon or a $K$ meson in proton and $K^\pm$ collisions, respectively) plus only pions. In particular, for an interaction of $E_{\gamma} < 2$ GeV the final-state will just include one or two extra pions, whereas at higher photon energies we take a multiplicity 
\beq
{\rm n_\pi} = {\rm Max}\left[ \,2,\; 2.3 \log_{10}(E_{\gamma}/ {\rm GeV}) \, \right].
\eeq
In these multi-pion diffractive collisions the leading baryon is a proton 2.2 times more frequently than a neutron, whereas pions appear in the three flavors with similar frequency. We assume equipartition of the initial energy among the final particles according to
\beq
    E_{i} = \frac{m_i}{\sum_j{m_j}}E_{h}.
\eeq
Finally, 
in a radiative photonuclear interaction the photon is sampled and treated as a real photon that is processed with the Monte Carlo simulator SIBYLL \cite{Riehn:2019jet}.

Let us discuss the effect of these EM processes by comparing the results in
{\it modified} runs of AIRES that include these EM processes with the results in {\it standard} runs. 
We will use the SIBYLL option and 0.1 GeV as the minimum 
photon energy, with a relative thinning of $10^{-4}$.
Each run contains 40.000 proton events from a zenith inclination $\theta_z=70^\circ$.
\begin{figure}
\begin{center}
\includegraphics[width=0.495\textwidth]{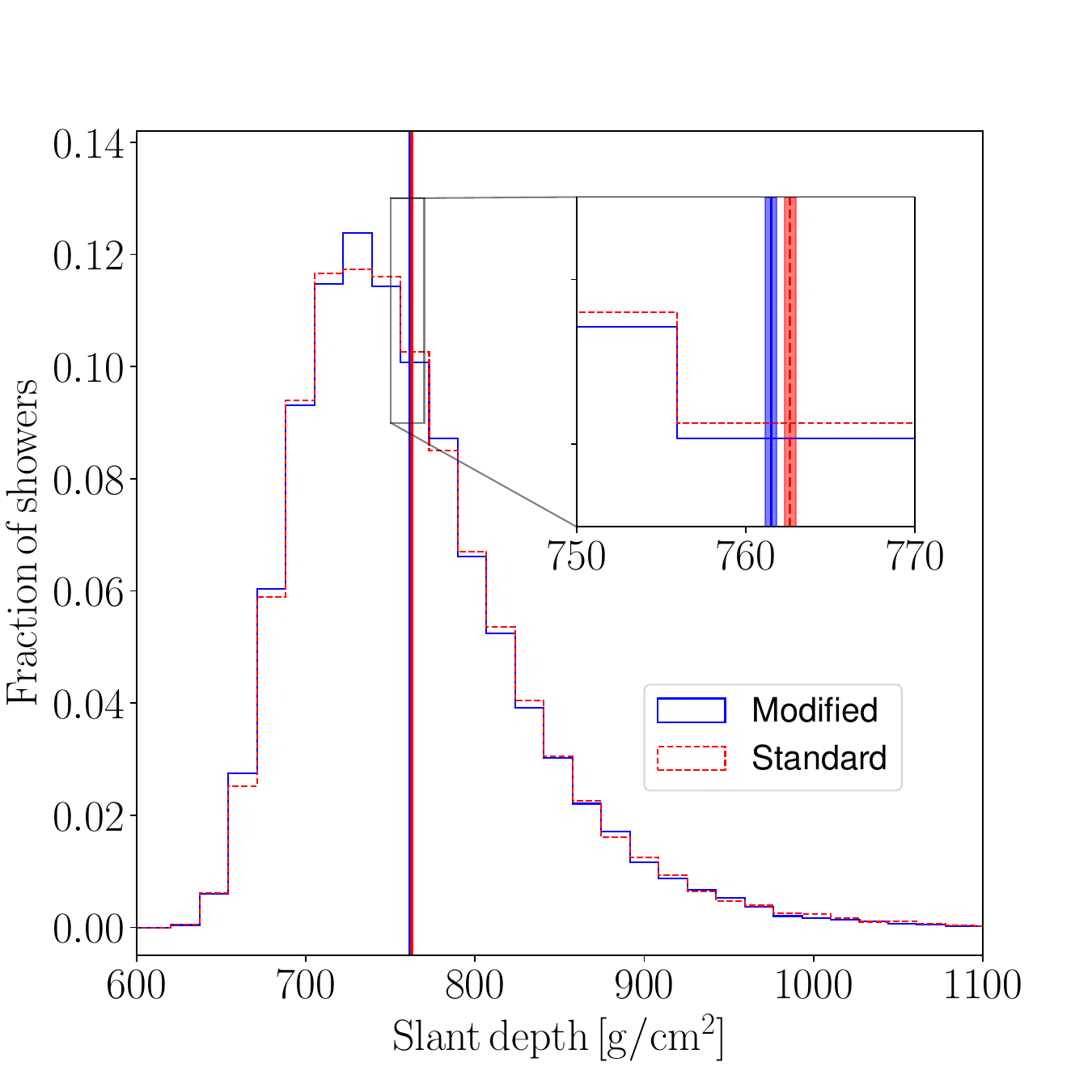}
\includegraphics[width=0.495\textwidth]{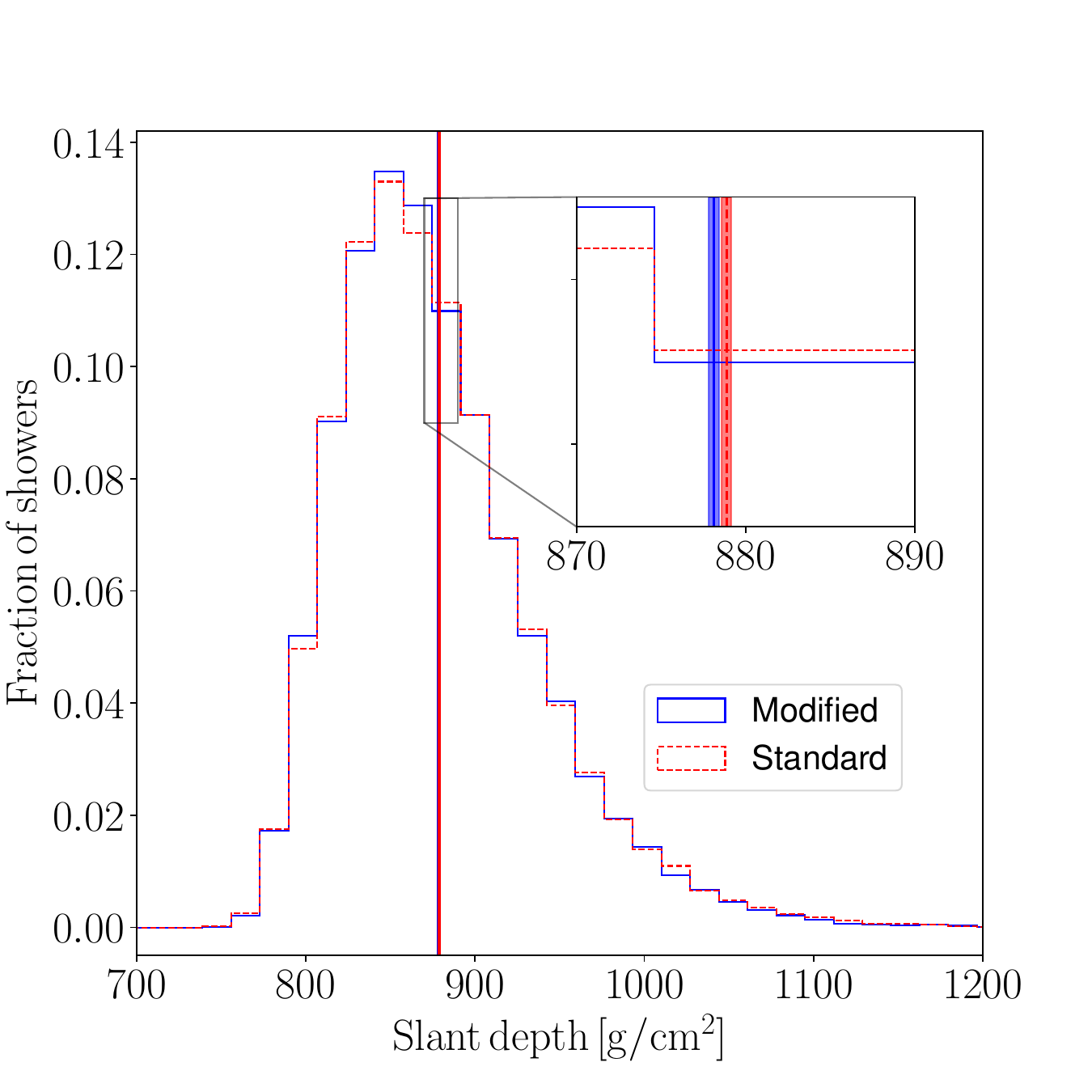}
\end{center}
\vspace{-0.5cm}
\caption{Distribution of $\Xm$ for 40.000 proton primaries of $E=10^9,\,10^{11}$ GeV (the bands indicate the statistical uncertainty).
\label{f5}}
\end{figure}
In Fig.~\ref{f5}, we plot the distribution of the shower maximum for primaries 
of $E=10^9,\,10^{11}$ GeV.
We observe that the inclusion of the EM processes increases the fraction of events with a small value of $\Xm$, but 
the effect on $\langle \Xm\rangle$ is just a reduction of $1.2$ g/cm$^2$ at $10^9$ GeV or of
$0.8$ g/cm$^2$ at $10^{11}$ GeV, with a statistical uncertainty of $0.3$ g/cm$^2$.
In Table \ref{t2}, we provide $\langle \Xm\rangle$ together with the value of the dispersion
$\Delta \Xm$, which in the modified run decreases by a 0.6\% at $10^{10}$ GeV.
\begin{table}[b!]
\centering
\begin{tabular}{ c | c c c}
 \hline
 \hline
$E$ [GeV] & $10^{9}$ & $10^{10}$ & $10^{11}$ \\
 \hline
$\langle \Xm^{\rm mod} \rangle \; [{\rm g/cm^2}]$ & $761.4$ & $819.5$ & $878.1$ \\
$\langle \Xm^{\rm st} \rangle \; [{\rm g/cm^2}]$ & $762.6$ & $820.4$ & $878.9$ \\
 \hline
$\Delta  \Xm^{\rm mod} \; [{\rm g/cm^2}]$ & $66.3$ & $62.3$ & $58.9$ \\
$\Delta \Xm^{\rm st} \; [{\rm g/cm^2}]$ & $67.3$ & $62.7$ & $59.8$ \\
 \hline
 \hline
\end{tabular}
\caption{Average value of $\Xm$ and $\Delta \Xm$ for 40.000 proton primaries of each energy.
\label{t2}}
\end{table}

\begin{figure}[t]
\begin{center}
\includegraphics[width=0.49\textwidth]{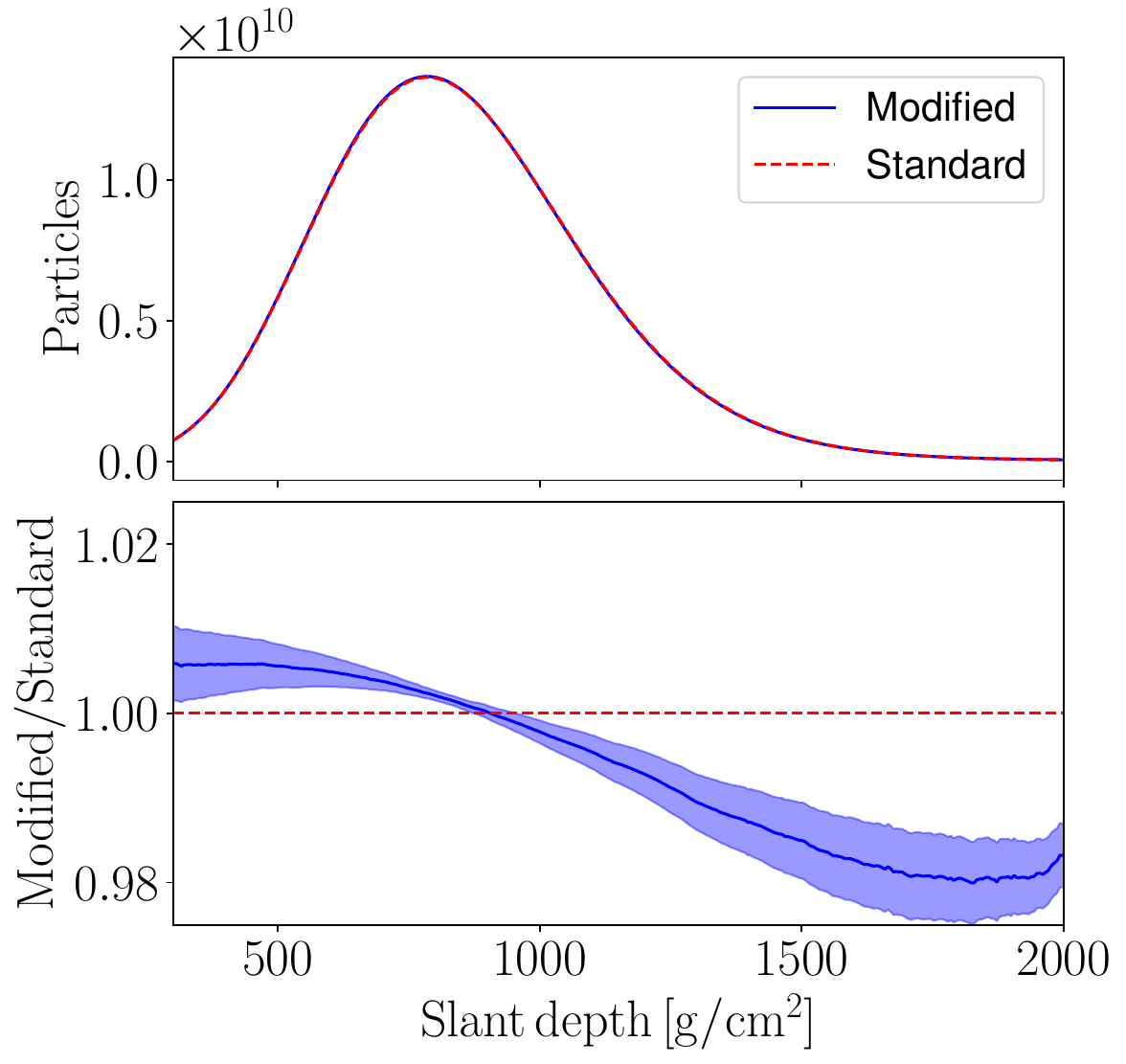}
\includegraphics[width=0.49\textwidth]{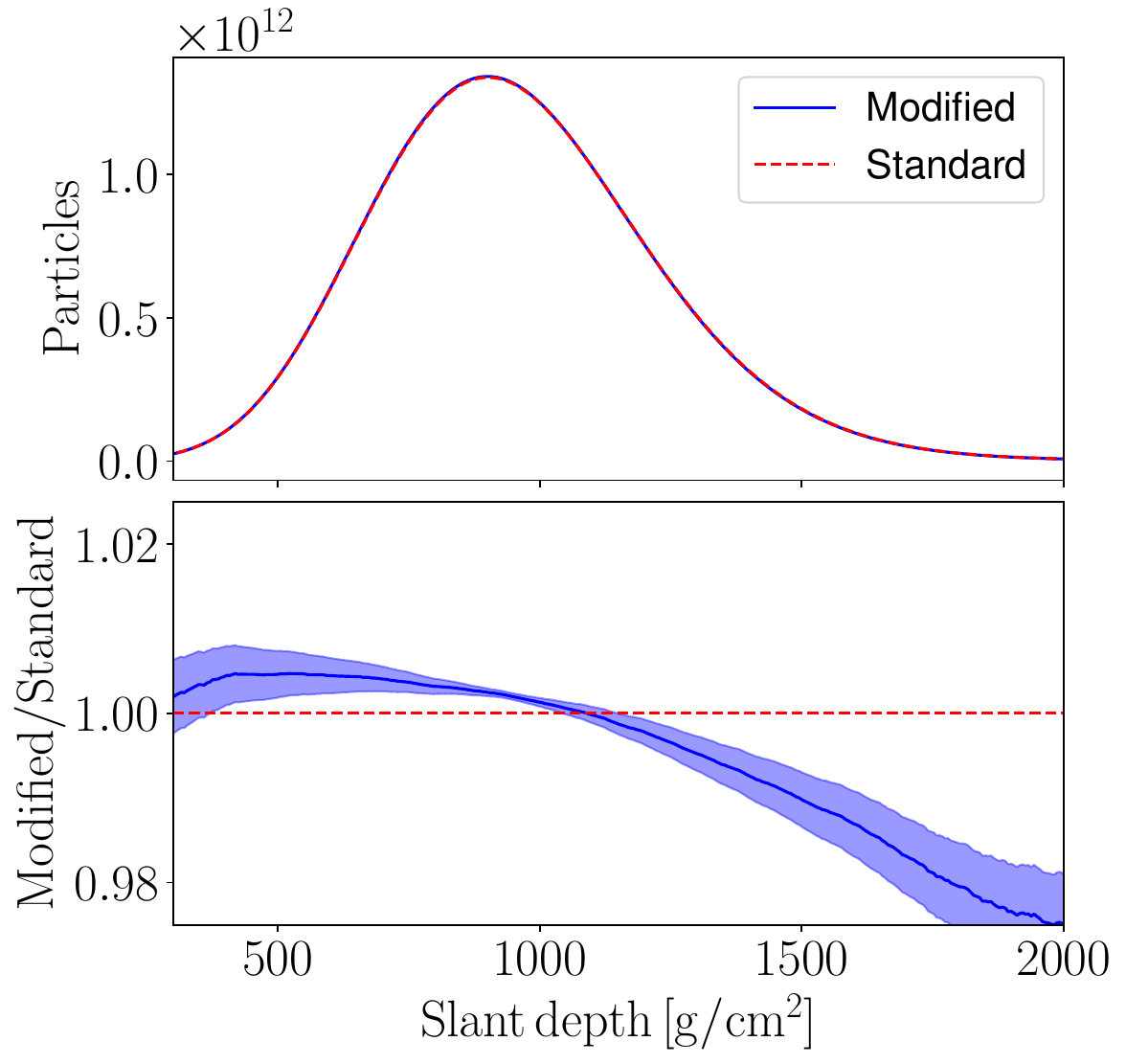}
\end{center}
\vspace{-0.5cm}
\caption{Total number of particles at different slant depths and relative difference between the
standard and modified runs for $10^9$ GeV (left) and $10^{11}$ GeV (right).
\label{f6}}
\end{figure}
\begin{figure}[t]
\begin{center}
\includegraphics[width=0.49\textwidth]{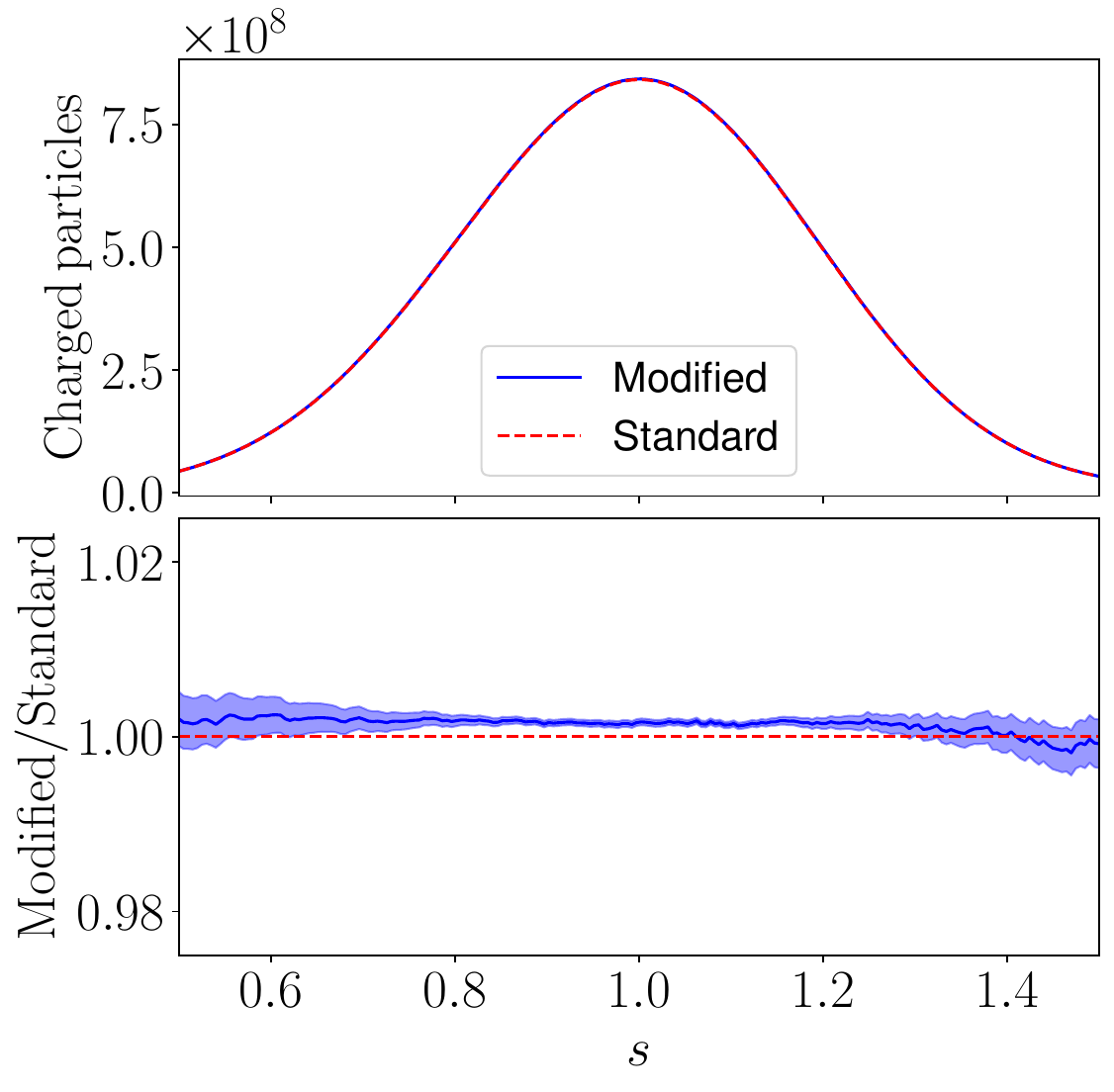}
\includegraphics[width=0.49\textwidth]{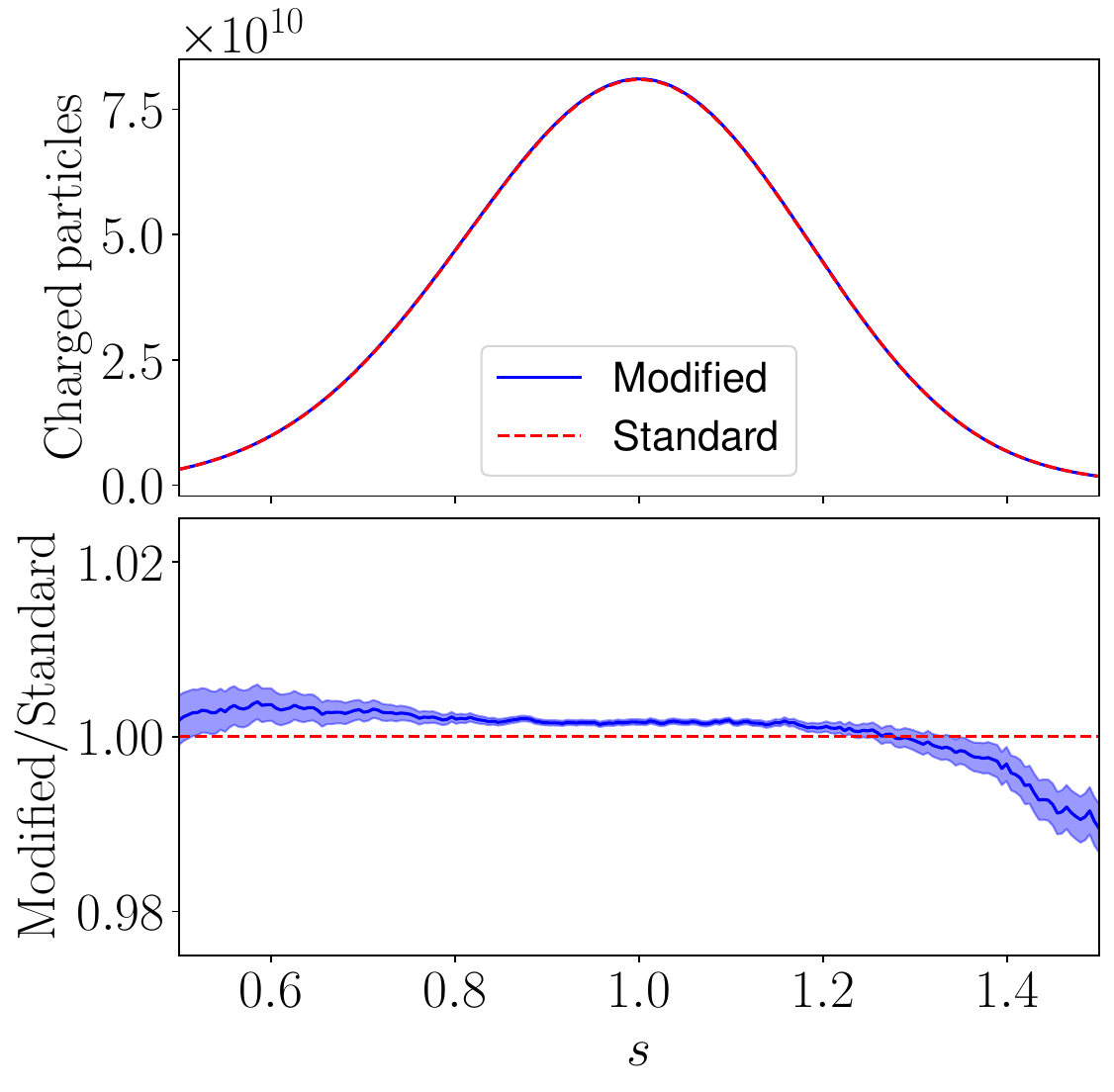}
\end{center}
\vspace{-0.5cm}
\caption{Number of charged particles in terms of the shower age $s$ as defined in Eq. (\ref{eq:def_s})
 for $10^9$ GeV (left) and $10^{11}$ GeV (right).
\label{f7}}
\end{figure}

The shift in $\langle \Xm\rangle$ implies a small reduction in the average number of particles at a given slant depth after $\Xm$, as we see Fig.~\ref{f6}. 
The effects are better understood if we center each shower at $\Xm$ and express the 
results in terms of the shower age $s$ \cite{Gora:2005sq}, 
\beq\label{eq:def_s}
s={3 X\over X+2\Xm}\,,
\eeq
with $s=1$ at $X=\Xm$
(see also \cite{Lipari:2008td} for a more accurate definition of the shower age).
In Fig.~\ref{f7}, we plot the number of charged particles for different values of $s$. We obtain that the effect of
the new interactions is an increase below 
1\% in the signal when the shower is young ($s\le 0.6$) and at the shower maximum, together with a decrease when the shower is old ($s\ge1.4$). Such variation, although not observable experimentally, seems
clear at the energies considered.

\begin{figure}[ht!]
\begin{center}
\includegraphics[width=0.49\textwidth]{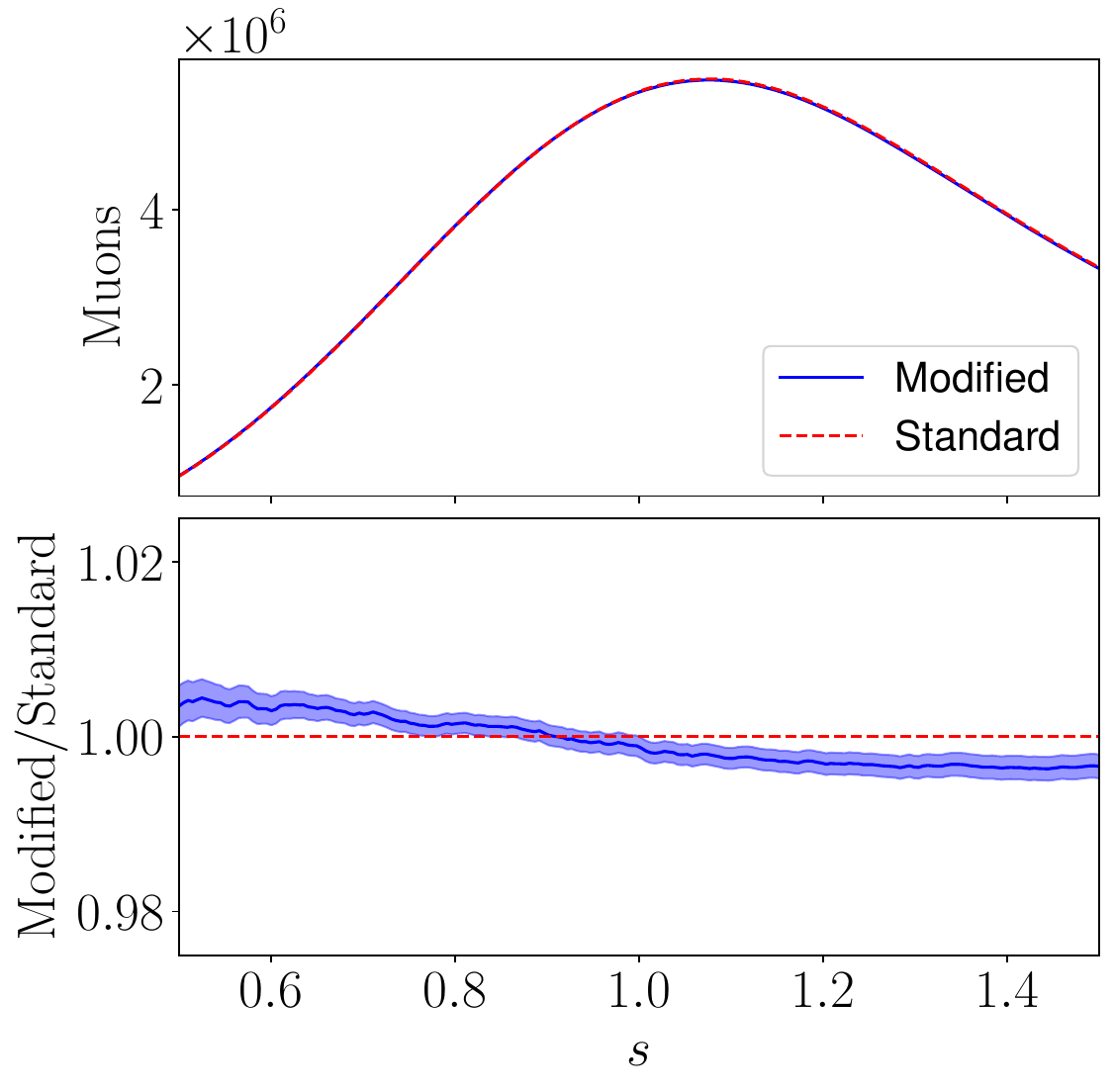}
\includegraphics[width=0.49\textwidth]{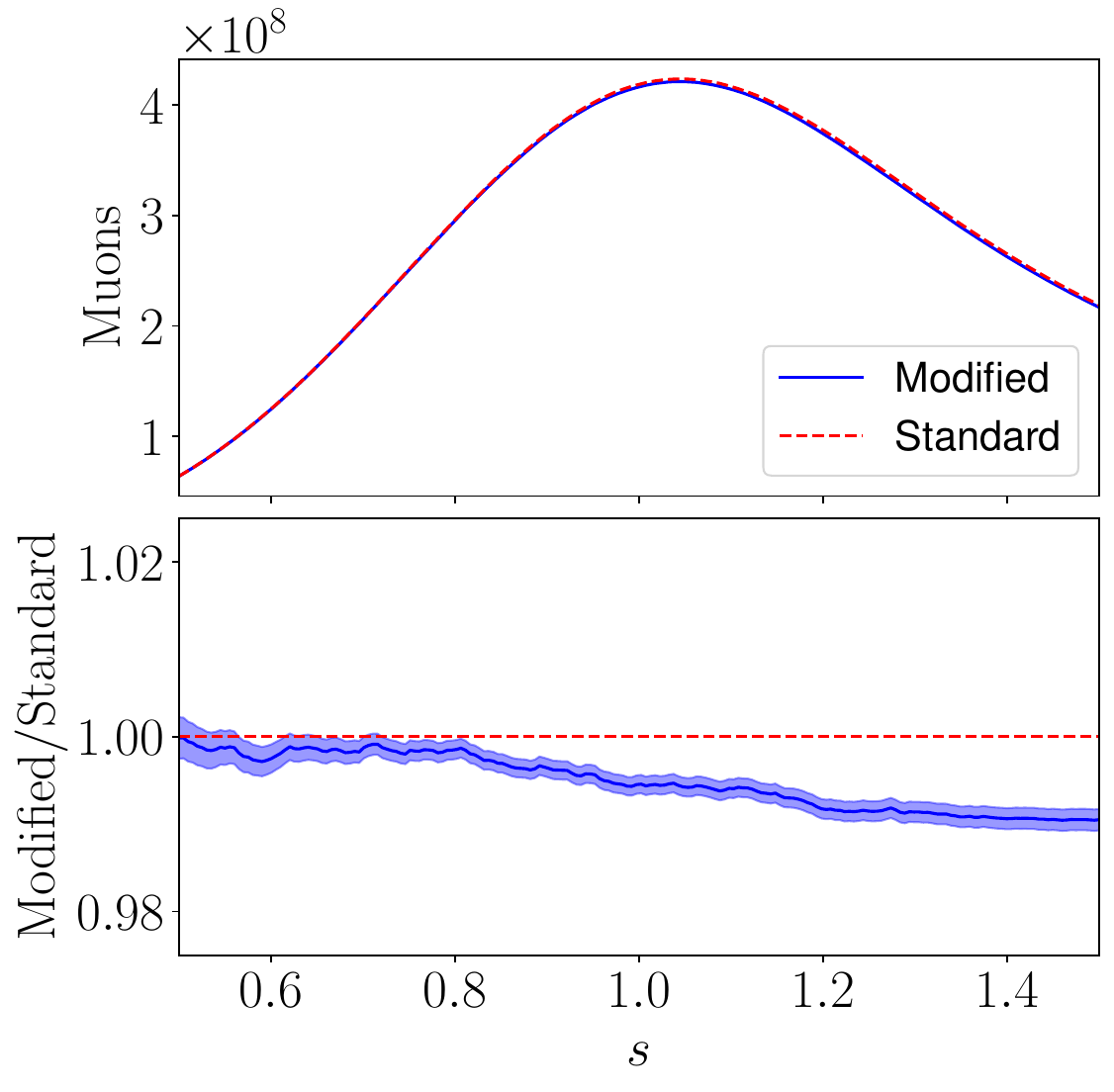}
\end{center}
\vspace{-0.5cm}
\caption{Number of muons in terms of the shower age $s$ as defined in Eq. (\ref{eq:def_s})
 for two different energies, $10^9$ GeV (left) and $10^{11}$ GeV (right).
\label{f8}}
\end{figure}
\begin{figure}[ht!]
\begin{center}
\includegraphics[width=0.49\textwidth]{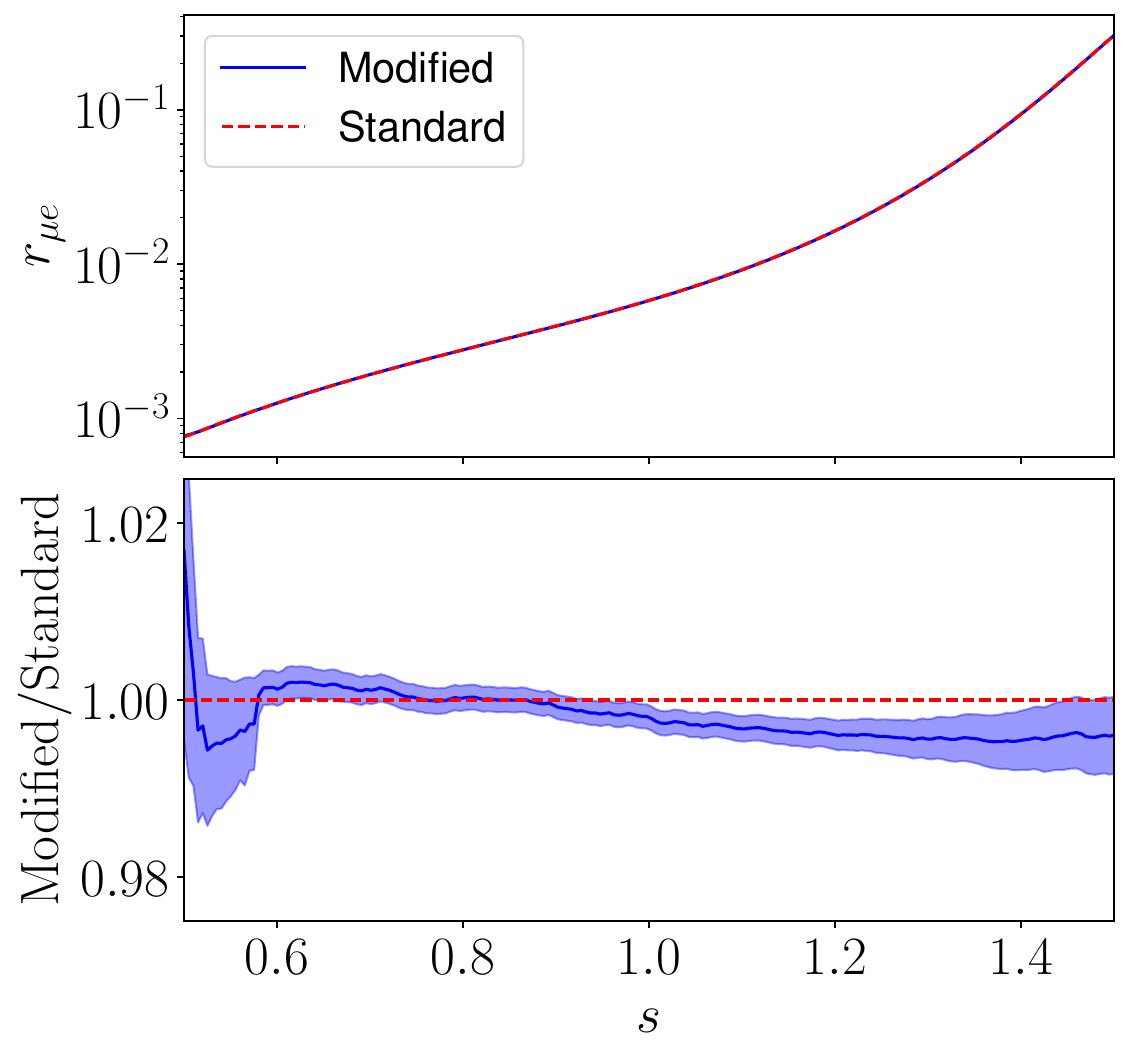}
\includegraphics[width=0.49\textwidth]{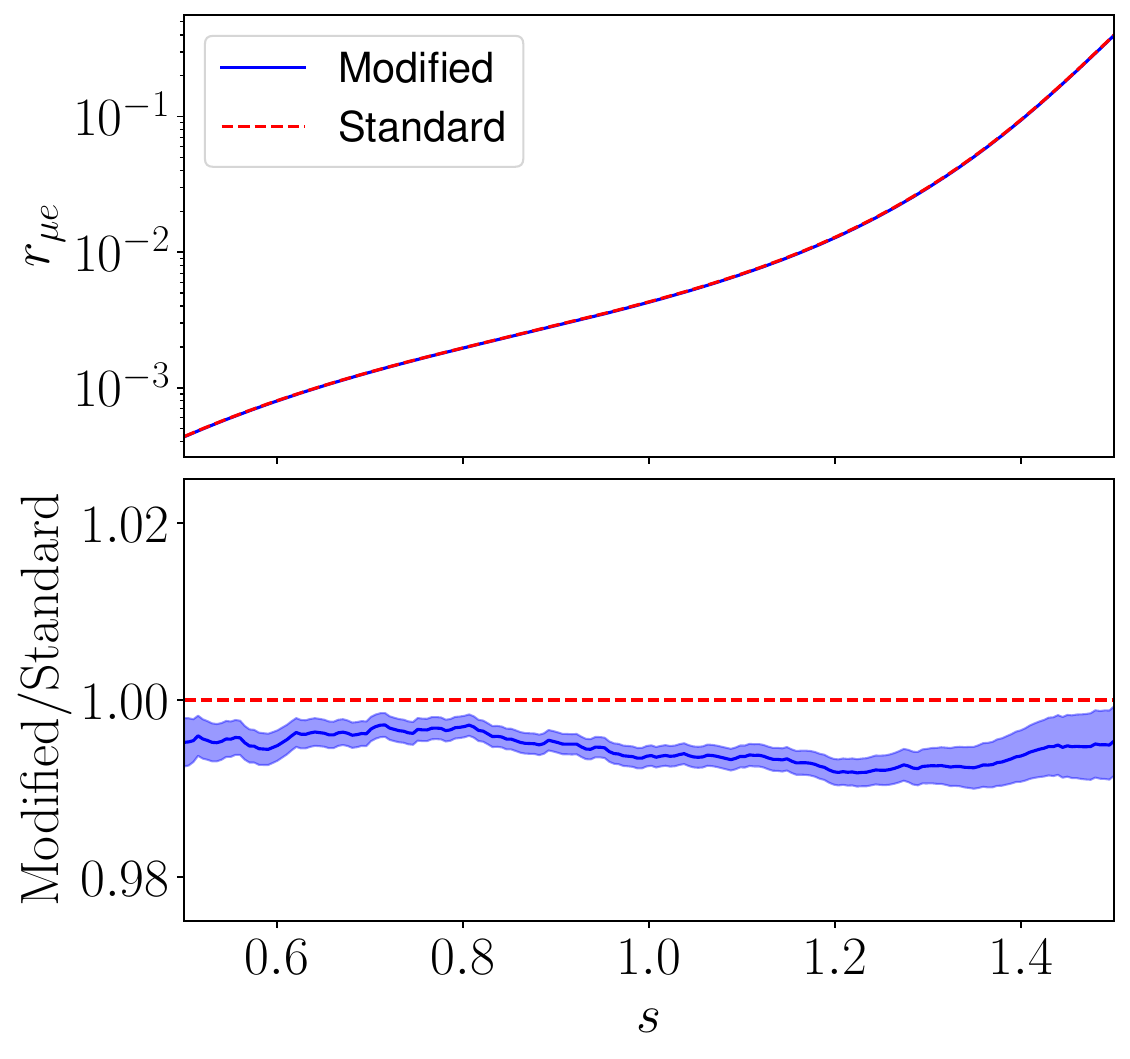}
\end{center}
\vspace{-0.5cm}
\caption{Muon to EM ratio $r_{\mu e}$ in terms of the shower age.
\label{f9}}
\end{figure}

The effect on the number of muons is illustrated in Fig.~\ref{f8}. We find that young showers include  
more muons (the excess is below 1\%) 
than in the average standard run, but as the shower develops the number of muons becomes 1\% smaller.
As a consequence, the showers evolve from $\Xm$ with a slightly poorer muon-to-electron ratio. This can be expressed in terms of 
$r_{\mu e}$ \cite{GarciaCanal:2016pev}, 
the ratio between the number of muons and the  energy of ($e^+ + e^-+\gamma$) in units of 500 MeV:
\beq
r_{\mu e}\equiv {n_\mu \over E_{e+\gamma}/(0.5\;{\rm GeV})}\,.
\eeq
We plot this observable in Fig.~\ref{f9}, where we appreciate a 1\% reduction at $s > 1$ due to the
new EM interactions. 

\section{Summary and discussion} 
The EM interactions of charged hadrons at very high energies 
are not included in current EAS simulators like AIRES or CORSIKA.
At these energies, the projectile may {\it break} when crossing the EM field of an air nucleus at relatively large
transverse distances (a diffractive collision), or
it may radiate a real photon (bremsstrahlung), or it may radiate a virtual photon that converts into a pair ($e^+ e^-$ emission) or a rho meson
(photonuclear collision). These ultra-peripheral processes have a longer interaction length 
than the hadronic ones, but we think that a precise estimate of their effect on EASs was long due. Here we have parametrized them 
and have then used AIRES to find the changes in $\Xm$ and in the muon or electron abundances at different slant depths that they introduce. We obtain 1\% corrections that, given the precision and the reduced statistics in EAS experiments, are far from being observable.

Despite the reduced size of these effects, they are significant (non-zero) and consistent. First of all, there is a slight excess (around 0.3\%) in the number of particles near $\Xm$; this excess becomes a deficit when the shower is old.
Second, the shower maximum is slightly shifted: the new interactions reduce in a few g/cm$^2$ the value of $\Xm$. This implies that at a given slant depth the showers are now a bit older. Third, old showers have less muons and a 1\% smaller muon-to-EM ratio. These effects just reflect {\it (i)} that the average shower develops {\it faster} due to the new collisions, and {\it (ii)} that the ratio of photons to charged pions introduced by these EM processes is larger than in hadronic collisions. 

Our results underline the consistency and the stability of current simulators under the type of processes considered.
Despite their sizeable cross section, ultra-peripheral collisions are events of low inelasticity, and their
inclusion in these simulators would improve their accuracy in just a 1\%. 

\section*{Acknowledgments}
This work was partially supported by the Spanish Ministry of Science, Innovation and Universities
(PID2019-107844GB-C21/AEI/10.13039/501100011033) and by Consejer\'\i a de Universidad, Investigaci\'on e Innovaci\'on de la Junta de Andaluc{\'\i}a / FEDER
(P18-FR-5057). 

\bibliography{ultra-peripheral}

\end{document}